\documentclass{INTERSPEECH2023}

\usepackage{multirow} 
\usepackage{booktabs} 
\usepackage{siunitx}  
\usepackage{float}
\usepackage{subfig}
\usepackage{amsmath,graphicx,url}
\usepackage{color,soul}
\usepackage{hyperref}

\interspeechcameraready

\title{Generating Multilingual Gender-Ambiguous Text-to-Speech Voices}
\name{
	\begin{tabular}{c}
		Konstantinos Markopoulos$^{1\star}$, 
		Georgia Maniati$^{1\star}$, \thanks{$^\star$Equal contribution}
		Georgios Vamvoukakis$^{1}$,
		Nikolaos Ellinas$^{1}$, \\
		Georgios Vardaxoglou$^{1}$,
		Panos Kakoulidis$^{1}$,
		Junkwang Oh$^{2}$,
		Gunu Jho$^{2}$,
		Inchul Hwang$^{2}$ \\
		Aimilios Chalamandaris$^{1}$,
		Pirros Tsiakoulis$^{1}$,
		Spyros Raptis$^{1}$
	\end{tabular}
}
\address{$^{1}$ Innoetics, Samsung Electronics, Greece \\
	$^{2}$ Mobile eXperience Business, Samsung Electronics, Republic of Korea}
\email{\{k.markop, g.maniati, s.raptis\} @samsung.com}

\begin{document}

\maketitle
 
\begin{abstract}
The gender of any voice user interface is a key element of its perceived identity. 
Recently, there has been increasing interest in interfaces where the gender is ambiguous rather than clearly identifying as female or male. 
This work addresses the task of generating novel gender-ambiguous TTS voices in a multi-speaker, multilingual setting. 
This is accomplished by efficiently sampling from a latent speaker embedding space using a proposed gender-aware method. 
Extensive objective and subjective evaluations clearly indicate that this method is able to efficiently generate a range of novel, diverse voices that are consistent and perceived as more gender-ambiguous than a baseline voice across all the languages examined. 
Interestingly, the gender perception is found to be robust across two demographic factors of the listeners: native language and gender. 
To our knowledge, this is the first systematic and validated approach that can reliably generate a variety of gender-ambiguous voices.
\end{abstract}


\section{Introduction}
\label{sec:intro}

Modern text-to-speech (TTS) synthesis systems are capable of producing high quality synthetic speech, which typically mimics the voice color and style of the speaker in the training data for single-speaker models, or one of the available speakers for models trained on a multi-speaker dataset.
The speaker identity representations learned by the TTS model in both cases correspond to a real person, who traditionally is a professional voice talent for commercial voices. 
The development of a new voice requires new high-quality recordings, a time-consuming and costly process.
Given the rising use of voice-enabled assistants, the availability of new human-sounding voices would offer a greater variety of options to UX designers in configuring TTS products to match the user needs. 

Greater user representation has been sought in the voice user interfaces in recent years. 
Female voices tend to no-longer be the sole or default option available to users, while the need for voice diversification and pluralism has risen.
Since the release of Q \cite{Qgenderless}, non-binary voices have gained attention in the market and the user communities, but relevant research on such voice generation is still very limited in the literature.

We investigate the generation of \textit{gender-ambiguous} voices that have the potential to offer greater flexibility to various TTS applications.
As the listeners are those who assign gender to a voice \cite{Sutton2020}, we use the term gender-ambiguous to refer to voices that may not be easily categorized by listeners in binary terms.
The scope of our work is thus targeted to investigating ways of generating new non-existent gender-ambiguous voices from multi-speaker binary data.

\subsection{Related work}
\label{ssec:subhead}

The speaker generation task has been recently introduced by Stanton \textit{et al} \cite{Stanton2022}.
In their work, they train a multi-speaker Tacotron model by using learnable speaker embeddings and create a speaker embedding prior to model the distribution over the speaker embedding space.
With this method, they are able to create new, natural and non-existent male and female voices. 
Subsequently, a similar approach has been presented in \cite{Bilinski2022}, where the speaker generation task is achieved through a flow-based TTS model \cite{Miao2020}. 

Relevant to gender-ambiguous voice generation, in specific, a recent lab report \cite{Yu2022} describes preliminary efforts of gender-free style transfer.
A gender style recognition model is trained to distinguish between male and female speaking style and is then used as a gender style encoder to extract gender embeddings from each utterance. 
These embeddings are used to train a Tacotron TTS model, and an attempt is made to calculate a gender-free embedding for use as input to the model during inference.
Gender information is also relevant when developing models that preserve the identity of a speaker. 
In this context, Stoidis and Cavallaro \cite{Stoidis2022} propose gender-ambiguous voice conversion using a GAN-based network and a MelGAN vocoder for privacy preservation purposes.

\subsection{Contribution}
\label{ssec:proposed}


In this work, we propose a method for generating multilingual gender-ambiguous voices in a zero-shot setting, i.e. from binary-gendered data only. 
While commercial approaches so far require the presence of one or several non-binary speakers in the training data, we generate novel voices by modeling the gender information in the speaker embedding space of a multilingual Tacotron and effectively sampling from it. 
Using objective metrics, we demonstrate that the generated voices are ambiguous with regards to their gender, diverse in voice color and consistent across languages.
Furthermore, we conduct subjective evaluations of naturalness and gender perception in 5 languages.
Our results testify that the novel voices generated with our method are superior to a baseline generated by simply averaging of the embeddings of all ground-truth speakers. 
Finally, we examine 2 demographic factors of our subjects, their language and gender, and find that the relative perception of the generated voices' gender is constant across both. 
To the best of our knowledge, this is the first systematic approach that can reliably generate a range of TTS voices, cross-lingually perceived as gender-ambiguous, that can meet diverse user requirements.

\section{Experimental Setup}
\label{sec:setup}


\subsection{Data}
\label{ssec:data}

We train our model on a multilingual multi-speaker dataset in US English (\textit{en}), Korean (\textit{ko}), Spanish (\textit{es}), German (\textit{de}) and French (\textit{fr}). It consists 
of 1391 hours and 1196 binary speakers, balanced as regards the speaker gender within each language. Both proprietary and open data are included, sampled at \SI{24}{\kilo\hertz} (Table~\ref{tab:data}).

\vspace{-12pt}
\begin{table}[h]
	\scriptsize
	\caption{Training dataset details}
	\vspace{-8pt}
	\label{tab:data}
	\centering
	\begin{tabular}{@{\hspace*{0mm}} l @{\hspace*{3mm}} l @{\hspace*{2mm}} l @{\hspace*{3mm}} S @{\hspace*{4mm}} S @{\hspace*{1mm}} S @{\hspace*{1mm}} S @{\hspace*{0mm}}}
			\toprule
			 &  & & & \multicolumn{3}{c}{\textbf{speakers}}\\
			\midrule
			\textbf{name} & \textbf{open} & \textbf{lng} & \textbf{hours} & \textbf{male} & \textbf{female} & \textbf{all}\\
			\midrule
			en96   & & en & 342 & 56 & 40 & 96\\
			LibriTTS \cite{Zen2019} & \checkmark & en & 163 & 457 & 421 & 878\\
			VCTK \cite{Veaux2017} & \checkmark & en & 25 & 46 & 62 & 108\\
			ko87  & & ko & 553 & 44 & 43 & 87\\
			es8   & & es &  96 & 4 &  4 & 8\\
			de9   & & de & 117 & 4 &  5 & 9\\
			fr10  & & fr & 95 &  4 &  6 & 10\\
			\bottomrule
		\end{tabular}
\end{table}
\vspace{-8pt}

We extract acoustic features that match the LPCNet vocoder \cite{Valin2019}: 20 Bark-scale cepstral coefficients (with additional 2 bands compared to LPCNet, due to higher sampling rate), as well as pitch period and pitch correlation. 
Text normalization and phonetization is performed with a TTS front-end module.

\subsection{Model architecture}
\label{ssec:model}

The TTS architecture is based on a non-attentive Tacotron \cite{Shen2020}. 
The model maps the input phoneme sequence into a sequence of acoustic feature frames used by the LPCNet vocoder \cite{Valin2019}.
The phoneme encoder converts input phonemes ${p} = [p^1, ..., p^N ]$ to learnable embedding vectors, which are processed by a 2-layer prenet and a CBHG stack from \cite{Wang2017} in order to produce the encoder representation ${e} = [e^1, ..., e^N ]$.
Speaker and language information are modeled by trainable embeddings ($s$ and $l$ respectively) produced by lookup tables, which are then broadcast concatenated to the phoneme encoder outputs.
The speaker embedding table $s$ has a fixed size of 256.
The phoneme-level sequence ${\hat{e}}=[{e}, {s}, {l}]$ is used as input to an LSTM based duration predictor, which is trained on phoneme durations extracted using forced alignment with Kaldi \cite{Povey2011}.
A Gaussian upsampler \cite{Shen2020} produces the frame-level representation from ${\hat{e}}$ based on the phoneme durations, which is then processed by the autoregressive LSTM decoder.
The final acoustic feature sequence is produced by a residual 5-layer convolutional postnet \cite{Shen2018}.
A reduction factor $r=4$ of the acoustic features is also used as it is shown to improve both quality and inference speed \cite{Ellinas2021}.
During the training phase, a linear adversarial classifier is trained to predict the speaker identity from the phoneme encoder outputs but with the addition of a gradient reversal layer, inducing disentanglement between phoneme and speaker information.
A figure with the model architecture can be found in the accompanying page\footnote{We encourage the reader to visit the accompanying page for additional figures and extensive samples at: https://innoetics.github.io/publications/gender-ambiguous/index.html}.

The Tacotron acoustic model required 23 hours of training on an NVIDIA RTX 4090 GPU with 24 GB RAM.  
We pair it with the LPCNet \cite{Valin2019} vocoder as adapted for reduced complexity by the parallel work of \cite{Vipperla2020}.

\section{Experiments and results}
\label{sec:experiments}

In this section, we present our experiments and results of gender-ambiguous voice generation\footnotemark[\value{footnote}]. 

\subsection{Embedding generation}
\label{sec:generation}

\begin{figure}[!t]
	\captionsetup[subfigure]{labelformat=empty}
	\centering
	\subfloat[\centering]{\includegraphics[width=3.8cm]{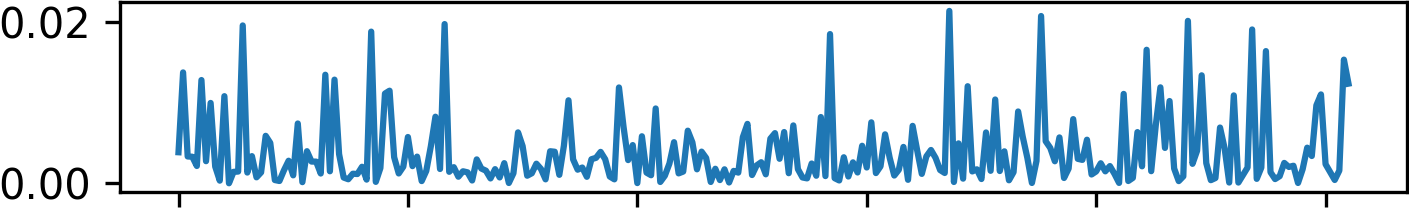}}%
	\qquad
	\subfloat[\centering]{\includegraphics[width=3.5cm]{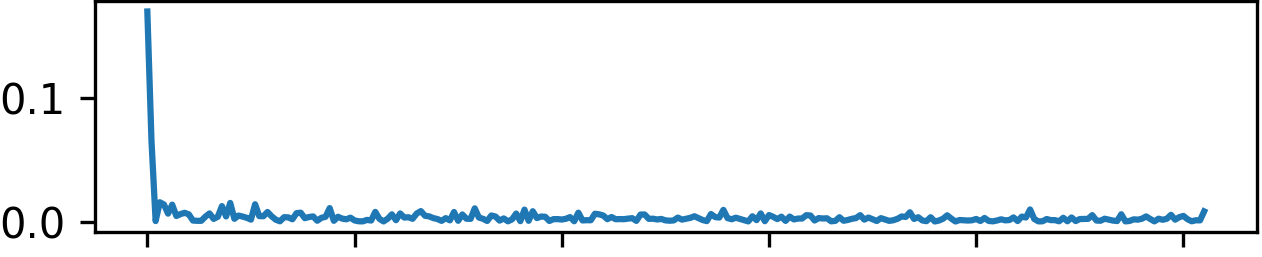}}%
	\vspace{-18pt}
	\caption{Correlation ratio of each dimension of the speaker embedding space to gender: (left) plain speaker embeddings, and (right) principal components of the speaker embeddings.}%
	\label{fig:correlations}
	\vspace*{-2mm}
\end{figure}

The 256-dimensional speaker embedding captures the speaker-specific characteristics of speech. 
Unavoidably, this includes all sources of acoustic variation, i.e. acoustic conditions, recording equipment, audio post-processing. 
Gender is one of the dominant sources of variation in speech \cite{Stanton2022}. 
To find the strength of the association of each speaker embedding dimension to gender, we use the correlation ratio,  $\eta$, calculated by dividing the weighted variance of the mean of each male/female category by the variance of all samples. 
As shown in Fig.~\ref{fig:correlations}, while gender information spreads across dimensions in the original speaker embedding space, performing principal component analysis (PCA) on the embedding factors the effect of gender into much fewer dimensions, notably the two most significant ones.
 
\begin{figure}[!t]
	\centering
	\includegraphics[width=6cm]{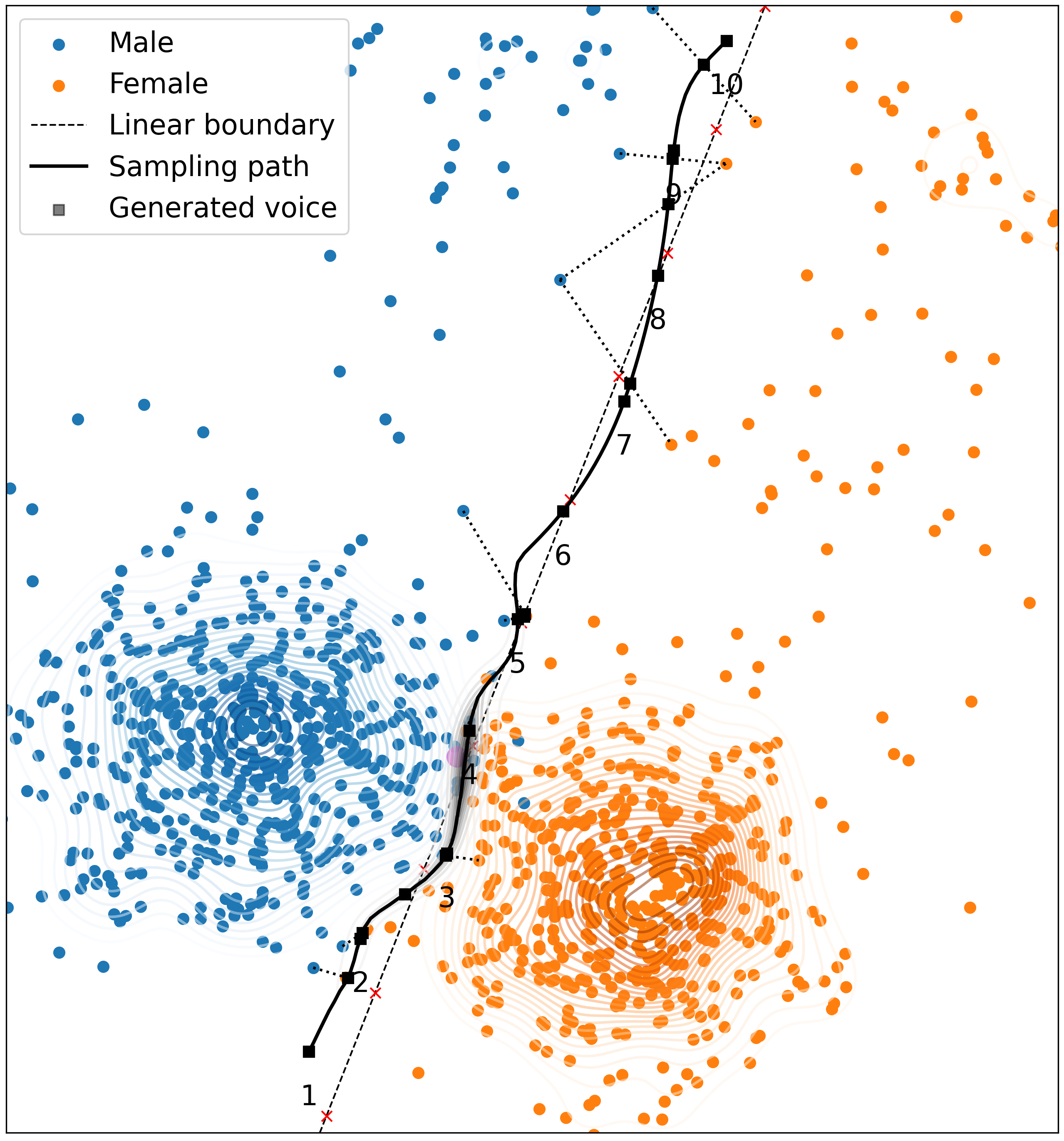}%
	\vspace*{-1mm}
	\caption{Density estimates for male (blue) and female (orange) speakers in the 2D PCA speaker embedding space. The derived pseudo-estimate for gender-ambiguous speakers density is shown in black contours. The thick black line is the sampling path along the local maxima of the density estimate. 10 suggested sampling points along the path are shown. The dashed line indicates the boundary of a linear male/female classifier.}%
	\label{fig:pca}
	\vspace*{-5mm}
\end{figure}

Fig. \ref{fig:pca} shows the projection of the embeddings of the speakers from the training dataset onto the first two principal components. 
As expected, the gender information appears very prominent and male/female speakers are almost linearly separable. 
Thus, a reasonable strategy for generating gender-ambiguous voices would be to appropriately sample from this 2D space at areas around the male/female boundary and then return to the original speaker embedding space through inverse PCA. 

Assuming that $P_m$ and $P_f$ are the density functions of male and female speakers in this space, we can calculate a (pseudo-) density estimate for the gender-ambiguous speakers that would assign high values to areas where $P_m$ and $P_f$ are similar (i.e. gender is ambiguous) and not very small (i.e. there are training speakers nearby) as:
\vspace*{-4mm}
\begin{align}
	P_{a(x,y)} = \frac{\min(P_{m(x,y)}, P_{f(x,y)})^2}{\max(P_{m(x,y)}, P_{f(x,y)})}
\end{align}

$P_m$ and $P_f$ have been estimated using kernel density estimation with Gaussian kernels, the haversine metric for pairwise distance computation, and an empirically selected bandwidth equal to $0.04$. 
They are shown in Fig.~\ref{fig:pca} by blue and orange contour lines, respectively. 
Black contour lines show $P_a$ calculated as above. 
Having a density function for gender-ambiguous voices provides a convenient way to generate different voices providing a variety of options. 
The thick black line in Fig.~\ref{fig:pca} follows the path along the local maxima of $P_a$. 
We can sample as many points as needed along that path. 
Fig.~\ref{fig:pca} shows 10 equidistantly sampled points but we only consider 5 of them in the following experiments (namely 1, 3, 5, 7, and 9).

After having sampled a point from the 2D PCA space using the above method, we use two alternative ways  to obtain appropriate values for the rest of the PCA dimensions so that we can transform it back to the original speaker embedding space in full 256 dimensions.
(a) A simple approach is to directly perform inverse PCA from just the 2D sampled point by using only the first two principal components. Practically, this is equivalent to assigning zero to all other PCA dimensions. 
As PCA centers the data in each of its dimensions, a zero value corresponds to the mean value in that dimension, thus offering a quite reasonable choice at the center of the data distribution as projected in that dimension.
(b) An alternative way is to pick the male and female speakers closest to the sampled point and then interpolate between their 256D PCA vectors to obtain plausible values for the rest of the dimensions. 
This interpolation is based on a weighted average:
\vspace*{-1.5mm}
\begin{align}
	E_a = \frac{(\frac{1}{d_m} \cdot E_m + \frac{1}{d_f} \cdot E_f)}{(\frac{1}{d_m} + \frac{1}{d_f})}
\end{align}
where $E_a$ is the derived speaker embedding for the gender-ambiguous speaker, $E_m$ and $E_f$ are the embeddings of the closest male and female speakers, and $d_m$ and $d_f$ are the distances of the first two dimensions of $E_m$ and $E_f$ from the sampled 2D point.

\begin{figure}[!t]
	\centering
	\includegraphics[width=7cm]{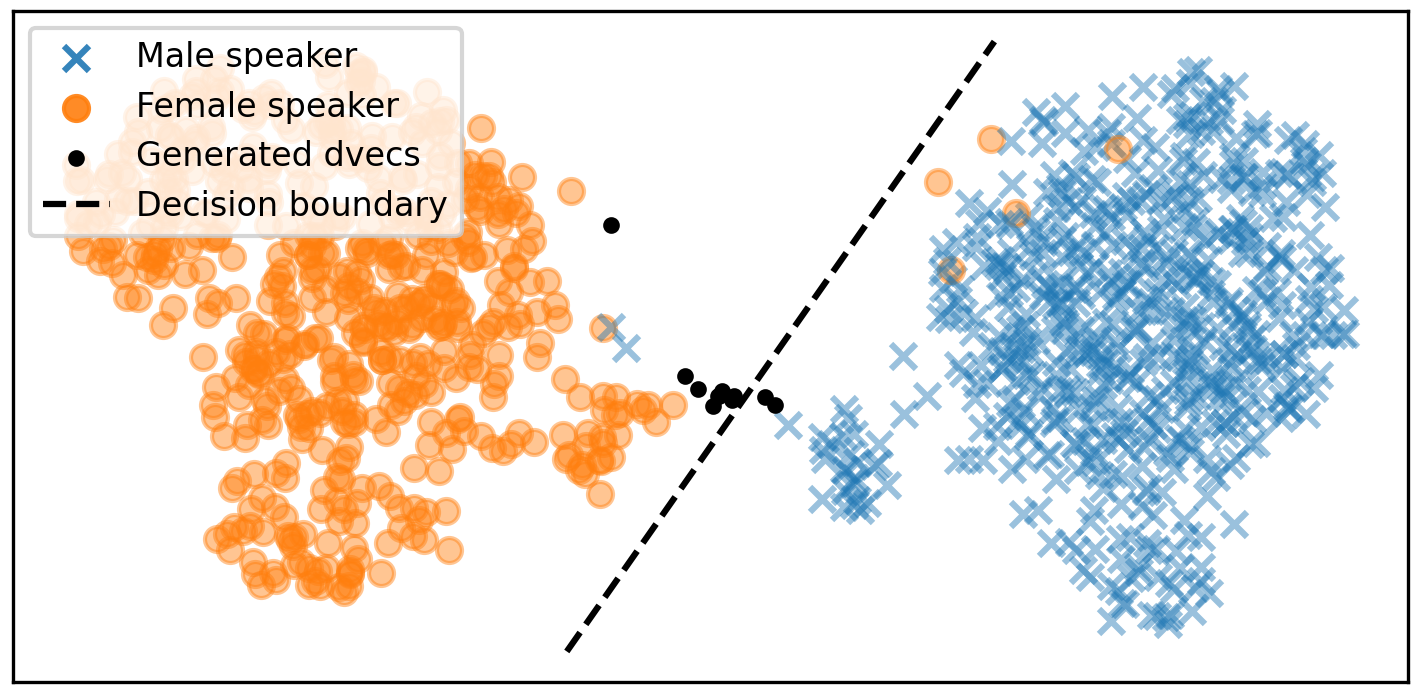}%
	\vspace*{-2mm}
	\caption{UMAP transformation of the d-vectors of the male (blue) and female (orange) speakers present in the dataset.}
	\label{fig:consist}
	\vspace*{-3mm}
\end{figure}

\begin{figure}[!t]
	\captionsetup[subfigure]{}
	\centering
	\subfloat[\centering]{\includegraphics[width=4.5cm]{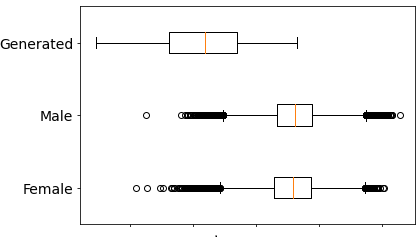}}%
	\qquad
	\subfloat[\centering]{\includegraphics[width=2.5cm]{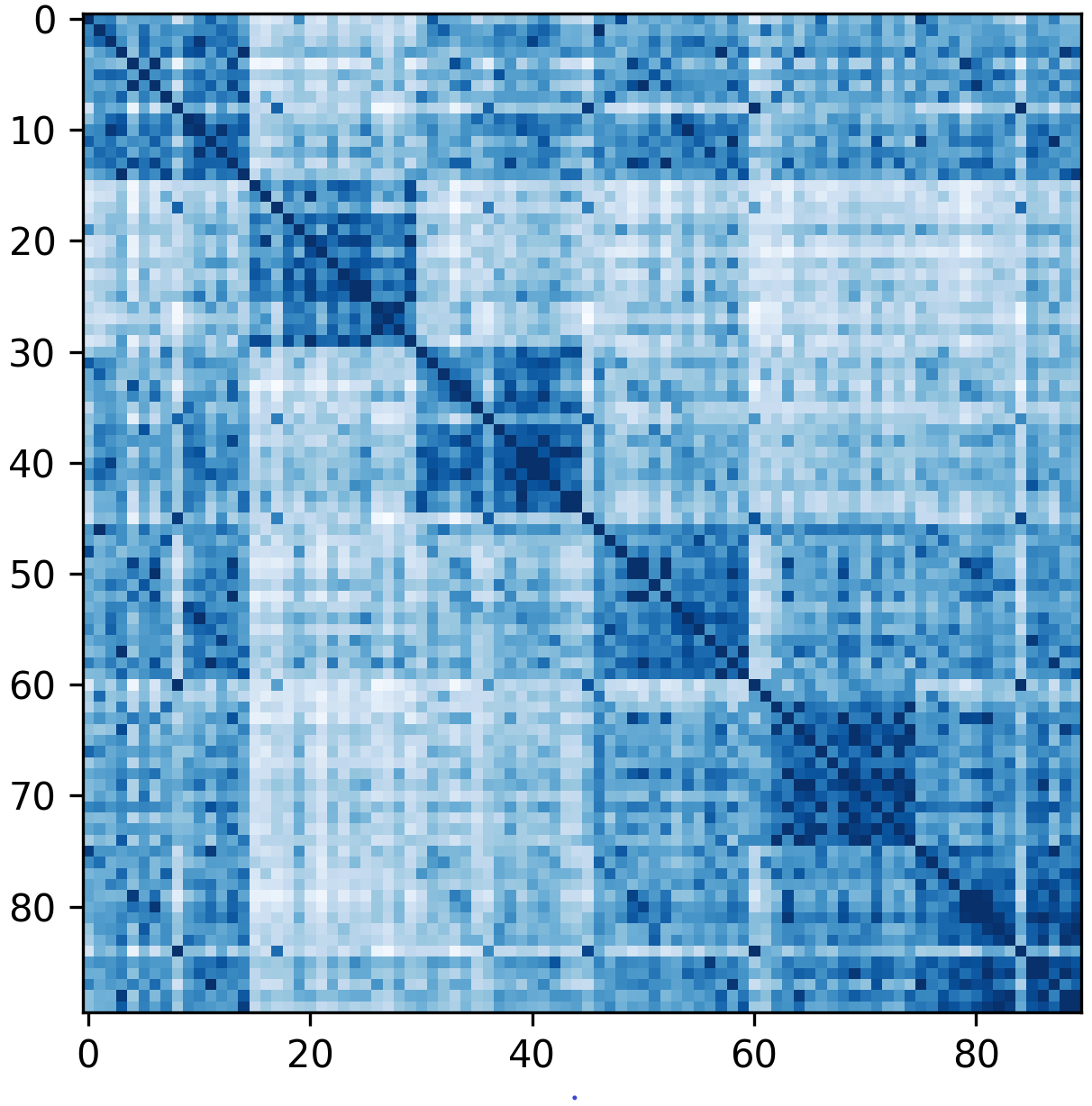}}%
	\vspace{-8pt}
	\caption{(a) Distribution of pairwise distances among (i) the male training voices, (ii) the female training voices, and (iii) the generated voices. (b) Distance matrix of 15 audio files from 6  generated voices (0, 1, 3, 5, 7, and 9).}
	\label{fig:utt-dists}
	\vspace*{-5mm}
\end{figure}

These two alternatives can generate novel speaker embeddings that share a common root in the 2D PCA space (which primarily associates with the gender), but are significantly differentiated in all their other aspects. 
Other methods could also be conceived. We speculate that no one of these methods will consistently deliver the “best” (most ambiguous) results in all cases, since their output depends on the actual data that the system is trained on, the shape of the resulting manifold in the higher dimensions, etc.
To assess the diversity of speakers provided by these methods, we generated embeddings with both, using the sampled points shown in Fig.~\ref{fig:pca}. 
We label the embeddings resulting from method (a) as $1$, $3$, $5$, $7$, $9$ and the ones from method (b) as $11$, $13$, $15$, $17$, $19$. 
These are complemented by the point “$0$” which corresponds to the mean of all speaker embeddings and is used as a baseline to compare against.

\begin{figure*}[t]
	\begin{minipage}[b]{1.0\linewidth}
		\centering
		\centerline{\includegraphics[width=19cm]{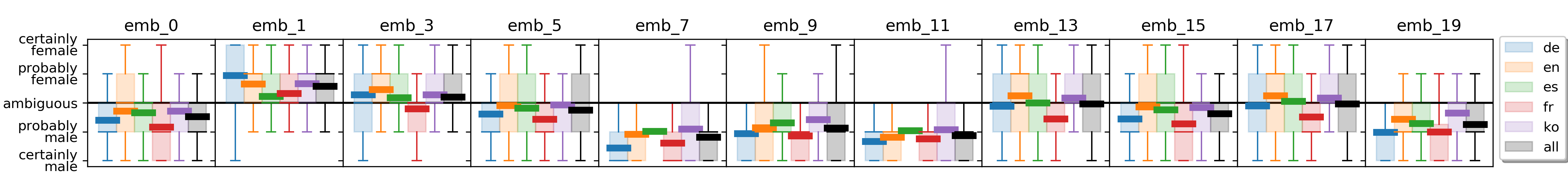}}
	\end{minipage}
	\vspace{-20pt}
	\caption{Cross-lingual MOS results of gender perception with 95\% confidence interval}
	\label{fig:gender}
	\vspace{-6pt}
\end{figure*}

\begin{table*}[t]
	\scriptsize
	\caption{Cross-lingual MOS results of naturalness with 95\% confidence interval}
	\vspace{-8pt}
	\label{tab:mos}
	\centering
	\begin{tabular}{@{\hspace*{0mm}} l @{\hspace*{1mm}} S @{\hspace*{-5mm}} S @{\hspace*{0mm}} S @{\hspace*{0.5mm}} S @{\hspace*{0mm}} S @{\hspace*{0.5mm}} S @{\hspace*{0mm}} S @{\hspace*{0.5mm}} S @{\hspace*{0mm}} S @{\hspace*{0.5mm}} S @{\hspace*{0mm}} S @{\hspace*{0.5mm}} S @{\hspace*{0mm}} S @{\hspace*{0.5mm}} S @{\hspace*{0mm}} S @{\hspace*{0.5mm}} S @{\hspace*{0mm}} S @{\hspace*{0.5mm}} S @{\hspace*{0mm}} S @{\hspace*{0.5mm}} S @{\hspace*{0mm}} S @{\hspace*{0.5mm}} S @{\hspace*{0mm}} S @{\hspace*{0.5mm}} S @{\hspace*{0mm}} S @{\hspace*{0mm}}}
		
		\toprule
		\multicolumn{2}{c}{\textbf{}} & \multicolumn{24}{c}{\textbf{Naturalness (MOS)}} \\
		
		\midrule
		\textbf{lng} & \textbf{}& \multicolumn{2}{c}{\textbf{emb\_0}} & \multicolumn{2}{c}{\textbf{emb\_1}} & \multicolumn{2}{c}{\textbf{emb\_3}} & \multicolumn{2}{c}{\textbf{emb\_5}} & \multicolumn{2}{c}{\textbf{emb\_7}} & \multicolumn{2}{c}{\textbf{emb\_9}} & \multicolumn{2}{c}{\textbf{emb\_11}} & \multicolumn{2}{c}{\textbf{emb\_13}} & \multicolumn{2}{c}{\textbf{emb\_15}} & \multicolumn{2}{c}{\textbf{emb\_17}} & \multicolumn{2}{c}{\textbf{emb\_19}} & \multicolumn{2}{c}{\textbf{gt}} \\
		
		\midrule
		\textbf{en}	&  & 3.31 & {\scriptsize${\pm}$0.13} & 3.11 & {\scriptsize${\pm}$0.14} & 2.83 & {\scriptsize${\pm}$0.15} & 3.07 & {\scriptsize${\pm}$0.14} & 3.55 & {\scriptsize${\pm}$0.13} & 3.19 & {\scriptsize${\pm}$0.14} & 3.53 & {\scriptsize${\pm}$0.13} & 3.29 & {\scriptsize${\pm}$0.12} & 3.38 & {\scriptsize${\pm}$0.13} & 3.31 & {\scriptsize${\pm}$0.13} & 3.48 & {\scriptsize${\pm}$0.13} & 4.70 & {\scriptsize${\pm}$0.06}\\
		\textbf{de}	&  & 3.51 & {\scriptsize${\pm}$0.18} & 3.08 & {\scriptsize${\pm}$0.20} & 2.91 & {\scriptsize${\pm}$0.19} & 3.22 & {\scriptsize${\pm}$0.19} & 3.78 & {\scriptsize${\pm}$0.18} & 3.41 & {\scriptsize${\pm}$0.18} & 3.27 & {\scriptsize${\pm}$0.19} & 3.43 & {\scriptsize${\pm}$0.18} & 3.50 & {\scriptsize${\pm}$0.18} & 3.43 & {\scriptsize${\pm}$0.18} & 3.65 & {\scriptsize${\pm}$0.17} & 4.89 & {\scriptsize${\pm}$0.06}\\
		\textbf{es}	&  & 3.24 & {\scriptsize${\pm}$0.11} & 2.28 & {\scriptsize${\pm}$0.13} & 2.30 & {\scriptsize${\pm}$0.12} & 2.98 & {\scriptsize${\pm}$0.12} & 3.33 & {\scriptsize${\pm}$0.12} & 3.02 & {\scriptsize${\pm}$0.11} & 3.00 & {\scriptsize${\pm}$0.11} & 3.03 & {\scriptsize${\pm}$0.11} & 3.18 & {\scriptsize${\pm}$0.11} & 3.32 & {\scriptsize${\pm}$0.11} & 3.50 & {\scriptsize${\pm}$0.11} & 4.81 & {\scriptsize${\pm}$0.05}\\
		\textbf{fr}	&  & 3.36 & {\scriptsize${\pm}$0.14} & 2.97 & {\scriptsize${\pm}$0.18} & 3.18 & {\scriptsize${\pm}$0.16} & 3.21 & {\scriptsize${\pm}$0.15} & 3.46 & {\scriptsize${\pm}$0.14} & 3.39 & {\scriptsize${\pm}$0.14} & 3.06 & {\scriptsize${\pm}$0.16} & 3.39 & {\scriptsize${\pm}$0.14} & 3.52 & {\scriptsize${\pm}$0.13} & 3.55 & {\scriptsize${\pm}$0.15} & 3.60 & {\scriptsize${\pm}$0.14} & 4.74 & {\scriptsize${\pm}$0.08}\\
		\textbf{ko}	&  & 4.17 & {\scriptsize${\pm}$0.13} & 3.91 & {\scriptsize${\pm}$0.13} & 3.81 & {\scriptsize${\pm}$0.13} & 4.10 & {\scriptsize${\pm}$0.14} & 4.08 & {\scriptsize${\pm}$0.14} & 3.99 & {\scriptsize${\pm}$0.13} & 4.02 & {\scriptsize${\pm}$0.14} & 4.26 & {\scriptsize${\pm}$0.13} & 4.24 & {\scriptsize${\pm}$0.13} & 4.27 & {\scriptsize${\pm}$0.14} & 4.15 & {\scriptsize${\pm}$0.14}& 4.91 & {\scriptsize${\pm}$0.05}\\
		
		
		\bottomrule
	\end{tabular}
	\vspace{-10pt}
\end{table*}

\subsection{Objective evaluation}
\label{sec:dvec}
We evaluate the generated embeddings objectively using \textit{d-vectors} \cite{wan2018,jia2018}, a widely used technique that derives a high-level representation of the characteristics of a voice and encodes it in a fixed size vector. 
We employ a pubic pre-trained d-vector encoder model.\footnote{https://github.com/resemble-ai/Resemblyzer}  
By design, similarity in the d-vectors space suggests similarity in the perceived voice timbre and vice versa. 
We rely on this to objectively assess various aspects of the generated voices, as discussed below.

(a) \textit{Gender ambiguity.} We consider the d-vector of a voice to be the average of the d-vectors of a few audios from that voice. 
To objectively assess the similarity of the generated voices to the training voices, we looked into their distances in the d-vectors space. 
Fig.~\ref{fig:consist} shows the UMAP of the d-vectors of the male, female and generated voices in 2D. 
Interestingly, the male and female speakers form two clearly separated clusters, while the generated voices a distinct third cluster. 
This figure reaffirms that gender is the major distinctive characteristic also in the d-vector space and verifies that the generated speakers indeed fall in between binary speakers.

(b) \textit{Voices diversity.} Capturing the geometric relations of a multidimensional dataset into two dimensions is generally not feasible, and approximate dimensionality reduction methods, such as UMAP or TSNE, are known to be poor at providing meaningful distances in the reduced space. 
So, the distances in Fig.~\ref{fig:consist} do not reflect how close or distinct the generated voices really are. 
To determine that objectively, we consider their distance in the d-vector space. 
Fig.~\ref{fig:utt-dists} (a) shows the distribution of pairwise distances among the male training voices, the female training voices, and the generated voices. 
Although the space of gender-ambiguous voices is much more constrained compared to male and female voices (as they need to reside at or near the gender boundary), the generated voices manage to retain a high degree of diversity.

(c) \textit{Voice consistency.} An important characteristic of a multilingual voice is its ability to maintain a consistent character across different languages. 
To objectively assess the consistency of our generated voices, we examine the distances of the d-vectors of audios synthesized with each of these voices in different languages. 
Fig.~\ref{fig:utt-dists} (b) shows the distance matrix of 15 audio files from 6 generated voices (0, 1, 3, 5, 7, and 9). 
Darker color indicates stronger similarity. 
Clearly, the d-vectors of utterances from each generated voice are significantly more similar to each other than those from other generated voices. 
As expected, the audio from the mean generated voice (voice 0) is somehow more similar to most of the generated voices.

\subsection{Subjective evaluation}
\label{sec:mos}

Our generated voices were assessed via mean opinion score (MOS) listening tests against naturalness and \textit{gender perception}.
Naturalness was evaluated on a Likert scale from ``1: very unnatural" to ``5: completely natural".
For gender perception, a custom test with five Likert-scale choices was created. 
Listeners were asked to rate how certain they are of the gender of the voice in each sample, from ``certainly male" to ``certainly female", with intermediate choices of ``probably male", ``neither male  nor female (ambiguous)" and ``probably female". 
The middle option has been our target during experimentation.
In this test type, we asked participants to disclose their gender in binary terms, if they wished.
Each test page comprised samples of 11 distinct generated voices (incl. baseline) as well as 1 validation and 1 ground truth sample (\textit{gt}) to allow post-filtering of results. 
In the gender perception test, the ratings of a page would be discarded in case a male gt sample was rated as ``probably/certainly female" and vice versa, while in naturalness the gt should be rated at least 3/5. 

Participants were asked to listen through headphones and be in a quiet setting. 
For European languages and English, we asked for 20 ratings via crowdsourcing on Amazon Mechanical Turk (AMT) \cite{Crowston2012} in the respective locales. 
For English, 3 locales were employed (US, GB, CA).
For Korean, for which AMT is not available, the evaluation was conducted on-site with 11 native speech experts. 
After assembling the results, we used the controls from our prior work \cite{Maniati2022} to exclude unreliable responses. 
Cross-lingually, 15432 ratings of 114 subjects were analyzed for gender perception and 14136 ratings of 102 subjects for naturalness.
Our listeners consisted of 58.4\% males, 31.7\% females and 9.9\% of undisclosed gender.

\begin{figure}[!t]
	\centering
	\vspace*{-2mm}
	\includegraphics[width=5cm]{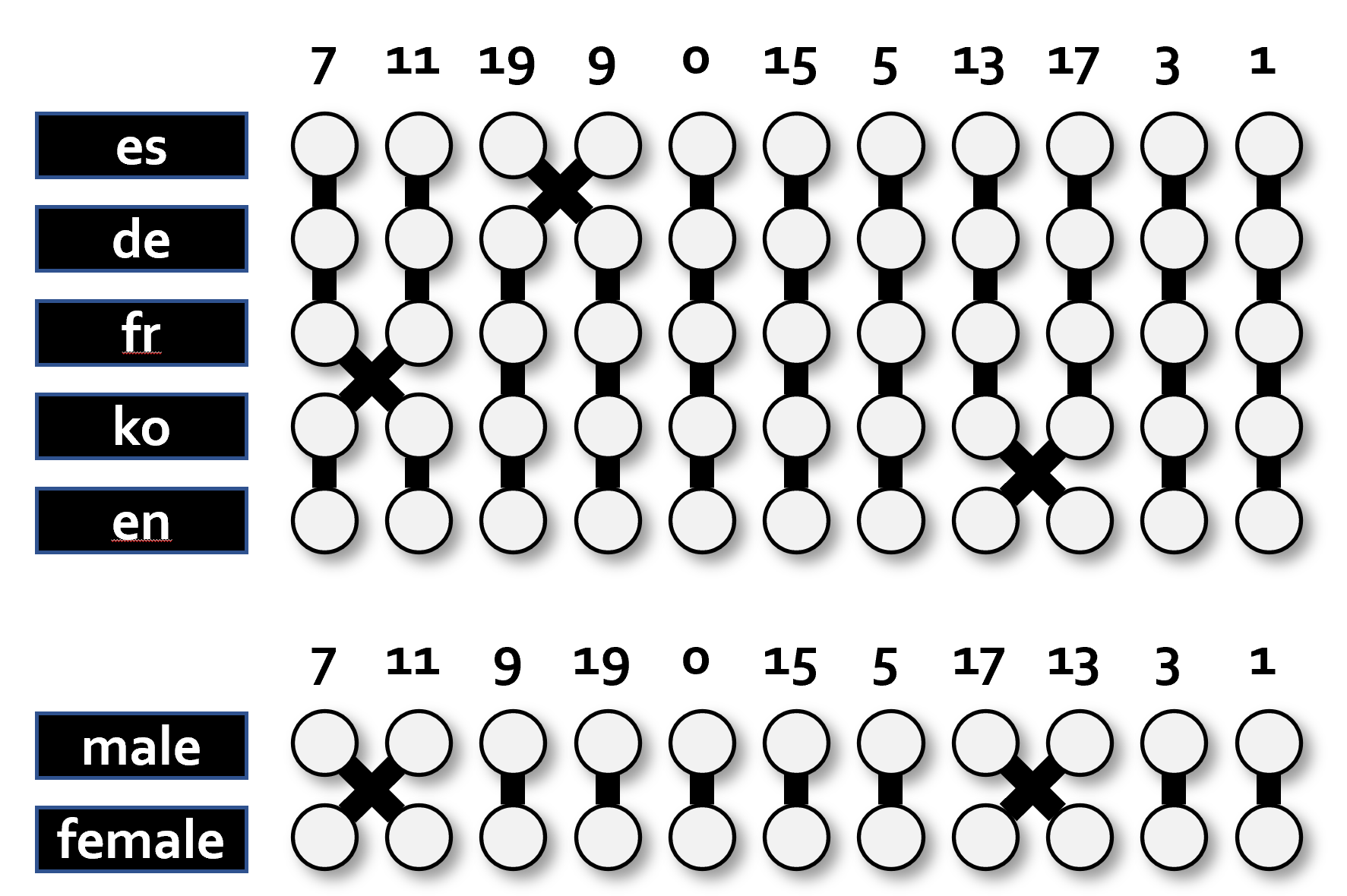}%
	\vspace*{-2mm}
	\caption{Ordering of generated voices from male to female, considering the listeners' demographics (language, gender).}
	\label{fig:order}
	\vspace*{-6mm}
\end{figure}

In Fig.~\ref{fig:gender}, the results of the cross-lingual gender perception tests are illustrated. 
This view comprises evaluations of all listeners, irrespective of their gender.
The baseline $0$ is close to gender-ambiguity but tends to be perceived as male, while our sampling approach has generated 5 voices rated as more ambiguous, depicted closer to the black horizontal line ($13$, $17$, $5$, $15$, $3$). 
Embeddings $7$, $11$ and $19$ are perceived as certainly/probably male, while $1$ as probably female.
It is interesting to note that the pairs $7$/$17$ and $1$/$11$ are gender-wise regarded different, although sharing the first 2 PCA dimensions.
This may be due to the fact that, although the first two PCA dimensions capture the largest part of the gender information, some part ``leaks" to other dimensions. 
So, depending on how specific values are obtained for those when a new speaker is generated, gender-dependent characteristics can leak back into the generated voice through higher dimensions, and certain embeddings may lean more towards a specific gender than others.
The overall gender-perception findings hold when considering the ratings of male or female listeners alone or grouped by the listeners' language (Fig~\ref{fig:order}). 
The ordering of systems from male to female is consistent, with few insignificant differences. 

Table~\ref{tab:mos} presents the MOS results of naturalness tests in all languages.
Overall, the most ambiguous voices ($13$, $17$) display similar or better naturalness (within CI) compared to the baseline $0$. 
In lack of natural, gender-ambiguous recordings to use as benchmarks in the test, we employed audio from the male/female training voices. 
This may have further amplified the difference between natural binary samples and synthesized gender-ambiguous samples, with which listeners were already unfamiliar, resulting in reduced scores for the latter. 
The assumption that naturalness assigned by naive listeners is influenced by voice gender plausibility can be observed in the higher MOS for male sounding voices ($7$, $19$) compared to ambiguous ones.
Notably higher are the MOS of Korean listeners who were experts, thus likely more comfortable in understanding the task.

\section{Conclusions}
We have proposed a method for generating zero-shot gender-ambiguous voices of satisfactory quality from binary-gendered data, by systematically sampling on the speaker embedding space of a multilingual multi-speaker non-attentive Tacotron model.
We have utilized objective ways including d-vectors, to validate our approach and showcase the diversity and consistency of our generated voices.
A large pool of subjective responses from cross-lingual experiments in 5 languages have been analyzed, which have shown that our method can generate voices that outperform a simple baseline, both in voice gender ambiguity and naturalness.
We have further examined the listeners' demographics and have found that neither their language nor their gender can significantly affect the perception of voice gender.
Instead, the relative perception of gender for the generated voices remains consistent across these factors. 

\bibliographystyle{IEEEtran}
\bibliography{mybib}

\begin{thebibliography}{10}
\providecommand{\url}[1]{#1}
\csname url@samestyle\endcsname
\providecommand{\newblock}{\relax}
\providecommand{\bibinfo}[2]{#2}
\providecommand{\BIBentrySTDinterwordspacing}{\spaceskip=0pt\relax}
\providecommand{\BIBentryALTinterwordstretchfactor}{4}
\providecommand{\BIBentryALTinterwordspacing}{\spaceskip=\fontdimen2\font plus
\BIBentryALTinterwordstretchfactor\fontdimen3\font minus
  \fontdimen4\font\relax}
\providecommand{\BIBforeignlanguage}[2]{{%
\expandafter\ifx\csname l@#1\endcsname\relax
\typeout{** WARNING: IEEEtran.bst: No hyphenation pattern has been}%
\typeout{** loaded for the language `#1'. Using the pattern for}%
\typeout{** the default language instead.}%
\else
\language=\csname l@#1\endcsname
\fi
#2}}
\providecommand{\BIBdecl}{\relax}
\BIBdecl

\bibitem{Qgenderless}
``{Meet Q: The first genderless voice - Full speech},''
  https://www.youtube.com/watch?v=lvv6zYOQqm0, 2019.

\bibitem{Sutton2020}
S.~J. Sutton, ``{Gender ambiguous, not genderless: Designing gender in voice
  user interfaces (VUIs) with sensitivity},'' in \emph{Proc. CUI}, 2020, pp.
  1--8.

\bibitem{Stanton2022}
D.~Stanton, M.~Shannon, S.~Mariooryad, R.~Skerry-Ryan, E.~Battenberg, T.~Bagby,
  and D.~Kao, ``Speaker generation,'' in \emph{Proc. ICASSP}, 2022, pp.
  7897--7901.

\bibitem{Bilinski2022}
P.~Bilinski, T.~Merritt, A.~Ezzerg, K.~Pokora, S.~Cygert, K.~Yanagisawa,
  R.~Barra-Chicote, and D.~Korzekwa, ``Creating new voices using normalizing
  flows,'' in \emph{Proc. INTERSPEECH}, 2022, pp. 2958--2962.

\bibitem{Miao2020}
C.~Miao, S.~Liang, M.~Chen, J.~Ma, S.~Wang, and J.~Xiao, ``{Flow-TTS: A
  non-autoregressive network for text to speech based on flow},'' in
  \emph{Proc. ICASSP}, 2020, pp. 7209--7213.

\bibitem{Yu2022}
C.~Yu, C.~Fu, R.~Chen, and A.~Tapus, ``First attempt of gender-free speech
  style transfer for genderless robot,'' in \emph{Proc. HRI}, 2022, pp.
  1110--1113.

\bibitem{Stoidis2022}
D.~Stoidis and A.~Cavallaro, ``Generating gender-ambiguous voices for
  privacy-preserving speech recognition,'' in \emph{Proc. INTERSPEECH}, 2022,
  pp. 4237--4241.

\bibitem{Zen2019}
H.~Zen, V.~Dang, R.~Clark, Y.~Zhang, R.~J. Weiss, Y.~Jia, Z.~Chen, and Y.~Wu,
  ``{LibriTTS: A corpus derived from LibriSpeech for text-to-speech},'' in
  \emph{Proc. INTERSPEECH}, pp. 1526--1530.

\bibitem{Veaux2017}
C.~Veaux, J.~Yamagishi, K.~MacDonald \emph{et~al.}, ``{CSTR VCTK corpus:
  English multi-speaker corpus for CSTR voice cloning toolkit},''
  \emph{University of Edinburgh. The Centre for Speech Technology Research
  (CSTR)}, 2017.

\bibitem{Valin2019}
J.-M. Valin and J.~Skoglund, ``{LPCNet: Improving neural speech synthesis
  through linear prediction},'' in \emph{Proc. ICASSP}, 2019, pp. 5891--5895.

\bibitem{Shen2020}
J.~Shen, Y.~Jia, M.~Chrzanowski, Y.~Zhang, I.~Elias, H.~Zen, and Y.~Wu,
  ``{Non-Attentive Tacotron: Robust and controllable neural TTS synthesis
  including unsupervised duration modeling},'' \emph{arXiv preprint
  arXiv:2010.04301}, 2020.

\bibitem{Wang2017}
Y.~Wang, R.~Skerry-Ryan, D.~Stanton, Y.~Wu, R.~J. Weiss, N.~Jaitly, Z.~Yang,
  Y.~Xiao, Z.~Chen, S.~Bengio \emph{et~al.}, ``{Tacotron: Towards end-to-end
  speech synthesis},'' in \emph{Proc. INTERSPEECH}, 2017, pp. 4006--4010.

\bibitem{Povey2011}
D.~Povey, A.~Ghoshal, G.~Boulianne, L.~Burget, O.~Glembek, N.~Goel,
  M.~Hannemann, P.~Motlicek, Y.~Qian, P.~Schwarz \emph{et~al.}, ``{The Kaldi
  speech recognition toolkit},'' in \emph{Proc. ASRU}, 2011, p. 1–4.

\bibitem{Shen2018}
J.~Shen, R.~Pang, R.~J. Weiss, M.~Schuster, N.~Jaitly, Z.~Yang, Z.~Chen,
  Y.~Zhang, Y.~Wang, R.~Skerrv-Ryan \emph{et~al.}, ``{Natural TTS synthesis by
  conditioning Wavenet on MEL spectrogram predictions},'' in \emph{Proc.
  ICASSP}, 2018, pp. 4779--4783.

\bibitem{Ellinas2021}
N.~Ellinas, G.~Vamvoukakis, K.~Markopoulos, A.~Chalamandaris, G.~Maniati,
  P.~Kakoulidis, S.~Raptis, J.~S. Sung, H.~Park, and P.~Tsiakoulis, ``High
  quality streaming speech synthesis with low, sentence-length-independent
  latency,'' in \emph{Proc. INTERSPEECH}, 2020, pp. 2022--2026.

\bibitem{Vipperla2020}
R.~Vipperla, S.~Park, K.~Choo, S.~Ishtiaq, K.~Min, S.~Bhattacharya,
  A.~Mehrotra, A.~G. C.~P. Ramos, and N.~D. Lane, ``{Bunched LPCNet: Vocoder
  for low-cost neural text-to-speech systems},'' in \emph{Proc. INTERSPEECH},
  2020, pp. 3565--3569.

\bibitem{wan2018}
L.~Wan, Q.~Wang, A.~Papir, and I.~L. Moreno, ``Generalized end-to-end loss for
  speaker verification,'' in \emph{Proc. ICASSP}, 2018, pp. 4879--4883.

\bibitem{jia2018}
Y.~Jia, Y.~Zhang, R.~Weiss, Q.~Wang, J.~Shen, F.~Ren, P.~Nguyen, R.~Pang,
  I.~Lopez~Moreno, Y.~Wu \emph{et~al.}, ``Transfer learning from speaker
  verification to multispeaker text-to-speech synthesis,'' \emph{Advances in
  neural information processing systems}, vol.~31, 2018.

\bibitem{Crowston2012}
K.~Crowston, ``{Amazon Mechanical Turk: A research tool for organizations and
  information systems scholars},'' in \emph{{Shaping the future of ICT
  research. Methods and approaches}}.\hskip 1em plus 0.5em minus 0.4em\relax
  Springer, 2012, pp. 210--221.

\bibitem{Maniati2022}
G.~Maniati, A.~Vioni, N.~Ellinas, K.~Nikitaras, K.~Klapsas, J.~S. Sung, G.~Jho,
  A.~Chalamandaris, and P.~Tsiakoulis, ``{SOMOS: The Samsung open MOS dataset
  for the evaluation of neural text-to-speech synthesis},'' in \emph{Proc.
  INTERSPEECH}, 2022, pp. 2388--2392.

\end{thebibliography}

\end{document}